\documentclass{emulateapj}

\def\kms{~\rm km~s^{-1}}

\def\IoTip{26.65} 
\def\eIoTip{0.09}

\def\tipdiffnobias{0.1} 

\def\MiTip{-3.97}  
\def\errMiTip{0.13}

\def\feh{-1.17}
\def\errfeh{0.25}  

\def\mMo{30.62}
\def\emMo{0.17}

\def\D{13.3}
\def\eD{1.0}

\def\Dlong{14.6} 

\def\Dold{13.8}  
\def\errDold{1.7} 

\def\EbvS{0.14}
\def\errEbvS{0.06}

\def\EbvSHR{0.046}

\def\tipACSlowred{26.57}
\def\tipACSlowredTrue{26.62} 
\def\errtipACSlowred{0.09}

\def\colortipACS{1.5}

\def\intextinc{0.16}
\def\errintextinc{0.11}

\def\intredd{0.09}
\def\errintredd{0.06}

\def\DWSninefive{28.8} 

\def\MvWSninefive{-15.5}

\def\newMassWetal{4\times 10^4}

\def\newLumIR{4 \times 10^{10}} 

\def\nULX{six}
\def\farcm{\hbox{$.\mkern-4mu^\prime$}}

\makeatother

\begin{document}

\title{A New Red Giant-based Distance Modulus of $\D$ Mpc to the Antennae
Galaxies and its Consequences\altaffilmark{1}}

\author{Ivo Saviane}

\affil{European Southern Observatory, A. de Cordova 3107, Santiago, Chile}

\email{isaviane@eso.org}

\author{Yazan Momany}

\affil{Osservatorio Astronomico di Padova, vicolo Osservatorio 5, 35122,
Padova, Italy}

\author{Gary S. Da Costa}

\affil{RSAA, Australian National University, Weston, ACT, 2611, Australia}

\author{R. Michael Rich}

\affil{Dep. of Physics and Astronomy, UCLA, 430 Portola Plaza, Los Angeles
CA, 90095-1547, USA}

\and{}

\author{John E. Hibbard}

\affil{National Radio Astronomy Observatory, 520 Edgemont Road, Charlottesville,
VA 22903, USA}

\altaffiltext{1}{Based on observations with the NASA/ESA \textit{Hubble
Space Telescope}, obtained at the Space Telescope Science Institute,
which is operated by the Association of Universities for Research
in Astronomy, Inc., (AURA), under NASA Contract NAS 5-26555.}

\begin{abstract}
The Antennae galaxies are the closest example of an ongoing major
galaxy merger, and thereby represent a unique laboratory for furthering
the understanding of the formation of exotic objects (e.g.,\ tidal
dwarf galaxies, ultra-luminous X-ray sources, super-stellar clusters,
etc). In a previous paper HST/WFPC2 observations were used to demonstrate
that the Antennae system might be at a distance considerably less
than that conventionally assumed in the literature. Here we report
new, much deeper HST/ACS imaging that resolves the composite stellar
populations, and most importantly, reveals a well-defined red giant
branch. The tip of this red giant branch (TRGB) is unambiguously detected
at $I_{\circ}^{{\rm TRGB}}=\IoTip\pm\eIoTip$ mag. Adopting the most
recent calibration of the luminosity of the TRGB then yields a distance
modulus for the Antennae of $(m-{M)}_{\circ}=\mMo\pm\emMo$ corresponding
to a distance of $\D\pm\eD$ Mpc. This is consistent with our earlier
result, once the different calibrations for the standard candle are
considered. We briefly discuss the implications of this now well determined
shorter distance. 
\end{abstract}


\section{Introduction}

The importance of having an accurate distance to the Antennae galaxies
(NGC~4038/4039) cannot be overstated. These galaxies are the nearest
major merger to the Milky Way, and as such, are taken as the archetypal
merger system. The proximity makes the Antennae of paramount importance
for studying mergers as a model for hierarchical galaxy formation,
as well as providing an ideal setting to explore the physical processes
in starforming galaxies. For example, the very idea of the formation
of exotic objects (e.g., \emph{ultra}-luminous X-ray sources, \emph{super}-stellar
clusters) in mergers owes much to studies of the Antennae. Indeed
there exists a huge amount of data collected over the years, from
ground and space missions, which might be precisely calibrated if
an accurate distance to the system were available. Recent examples
include: GALEX observations (Hibbard et al. \citeyear{hibbard_etal05}),
near-IR WIRC imaging (Brandl et al. \citeyear{brandl_etal05}); Chandra
ACIS-S (Baldi et al. \citeyear{baldi_etal06}); STIS data (Whitmore
et al. \citeyear{whitmore_etal05}); VLT/ISAAC data (Mengel et al.
\citeyear{mengel_etal05}); and VLT/VIMOS data (Bastian et al. \citeyear{bastian_etal06}).
We also note the recent Type Ic supernova in the Antennae (SN 2004gt,
IAUC 8456; Gal-Yam et al. \citeyear{gal-yam_etal05}; Maund et al.
\citeyear{maund_etal05}), but the nature of its progenitor remains
a subject of debate in part due to the unresolved issue of the distance.
Indeed, all basic physical quantities that enter the models (linear
distances, masses, and luminosities), and hence the model predictions,
depend on the adopted distance.

The most widely quoted distance to the Antennae is $\sim20$~Mpc,
based on the recession velocity of the system corrected with the flow
model of Tonry et al. (\citeyear{tonry_etal00}). However values up
to $\sim$30 Mpc are sometimes quoted (e.g.,\ Fabbiano et al. \citeyear{fabbiano_etal01},
Zezas \& Fabbiano \citeyear{zezas_fabbiano02}). On the other hand,
significantly shorter distance measurements were provided in Rubin
et al. (\citeyear{rubin_etal70}) and Saviane, Hibbard, \& Rich (\citeyear{shr04};
hereafter SHR04). The first of the two studies suggested a distance
between $6$ and $13$ Mpc. The large distance uncertainty is due
to the relatively low precision of their adopted distance indicators
(the size of \ion{H}{2} regions, the brightest stars, and the
1921 supernova), as well as to an unknown contribution from internal
absorption. Yet they were the first to suggest a distance shorter
than that predicted by the redshift.

The second study offered a much more accurate measurement of the distance
to the Antennae system based on a standard candle. SHR04 used HST/WFPC2
images to investigate the stellar populations of the candidate tidal
dwarf galaxy that lies at and beyond the tip of the southern tidal
tail (cf.\ Schweizer \citeyear{schweizer78}, Mirabel, Dottori, \&
Lutz \citeyear{mdl92}). The SHR04 color-magnitude diagram (CMD) revealed
an apparently spatially extended population of old (age $\geq$ 2
Gyr) red giants, underlying the more concentrated star-forming regions.
The detection of such a red giant branch (RGB) population was unexpected,
since for an assumed distance of at least 19-20 Mpc, such a population
would be fainter than $I\simeq27.5$, beyond the detection limit of
the SHR04 photometry. Given that all other possible interpretations
of this faint red stellar population were unsatisfactory, SHR04 assumed
that the tip of the RGB (TRGB) had indeed been reached. Stellar populations
older than $\sim1$--$2$~Gyr develop a well populated red giant
branch. The luminosity of the brightest red giants is limited by the
ignition of helium in the core, under degenerate conditions, the so-called
helium flash. This luminosity corresponds to $M_{I}\sim-4$ and is
known as the red giant branch tip. With $I_{\circ}^{{\rm TRGB}}\simeq26.5$,
and assuming an absolute luminosity of $M_{I}=-4.2$ (Carretta et
al. \citeyear{carretta_etal00}), a distance to the Antennae of $\sim$13.8
Mpc follows, a value significantly smaller than those previously assumed.
Admittedly, in SHR04 the TRGB was detected near the completeness limit
of the imaging, where large errors and crowding affect the precision
of the photometry. It is therefore reasonable to require confirmation
of this shorter distance before it is universally accepted.

Such confirmation was the motivation for the new HST/ACS observations
presented in this paper, which reach significantly fainter magnitudes
than the WFPC2 data. We use this ACS photometry to derive a firm distance
for the Antennae, finding a value consistent with that given in SHR04.

\section{Observations and reductions \label{sec:Observations-and-reductions}}

\begin{figure}
\begin{centering}
\includegraphics[width=1\textwidth]{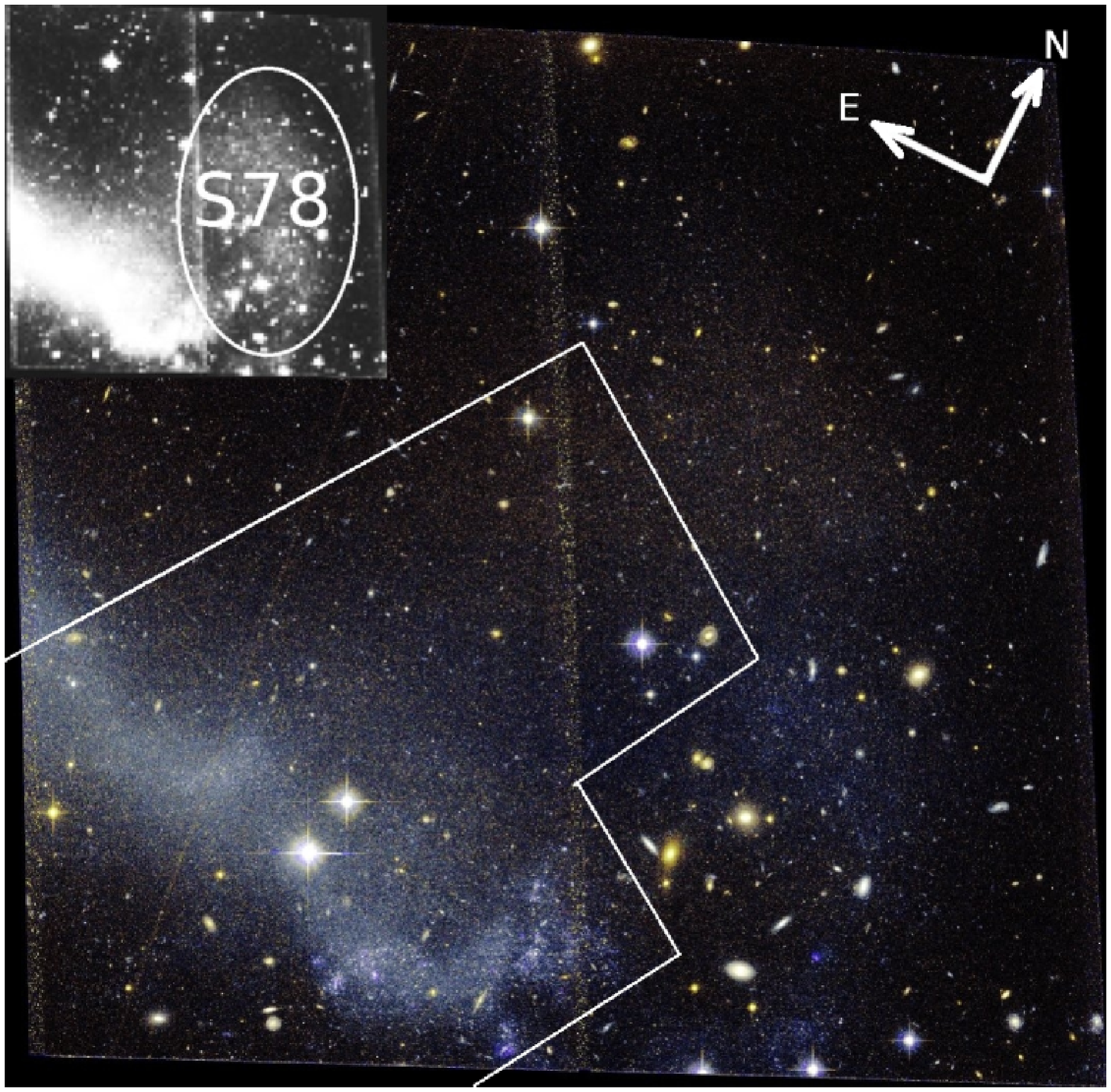} 
\par\end{centering}

\caption{A false color-image of the tip of the southern tidal tail of the Antennae
galaxies, constructed from the HST $F606W$ and $F814W$ images. The
star-forming regions are evident but a large fraction of the frame
is evidently free of active star formation. In the inset the low-luminosity
regions are enhanced to show the {tidal feature} identified
by Schweizer (\citeyear{schweizer78}) beyond the tip of the tail.
The {feature} is marked by the ellipse and the label `S78'.
The outline shows the location of the WFPC2 field studied in SHR04.
The field of view is $3\farcm5\times3\farcm5$.\label{f_figura1} }

\end{figure}

ACS/WFC observations of the NGC~4038 southern tidal tail (GO 10580,
P.I. Saviane) were obtained on December 30, 2005, and January 3 and
4, 2006; they were centered at\\
 ($\alpha,\delta$)=($12$:$01$:$26.511,-18$:$59$:$21.35$) (see
Fig.~\ref{f_figura1}). Of the 7 awarded orbits, 4 were dedicated
to deep observations through the $F606W$ and 3 through the $F814W$
filters, for a total integration time of 10870~s and 8136~s, respectively.
No cosmic ray split was applied, but the 7 deep exposures were dithered
properly so that the frames, when combined via the drizzle process,
filled the inter-chip gap and allowed removal of the cosmic-ray contamination.

\begin{figure}
\begin{centering}
\includegraphics[width=0.6\textwidth]{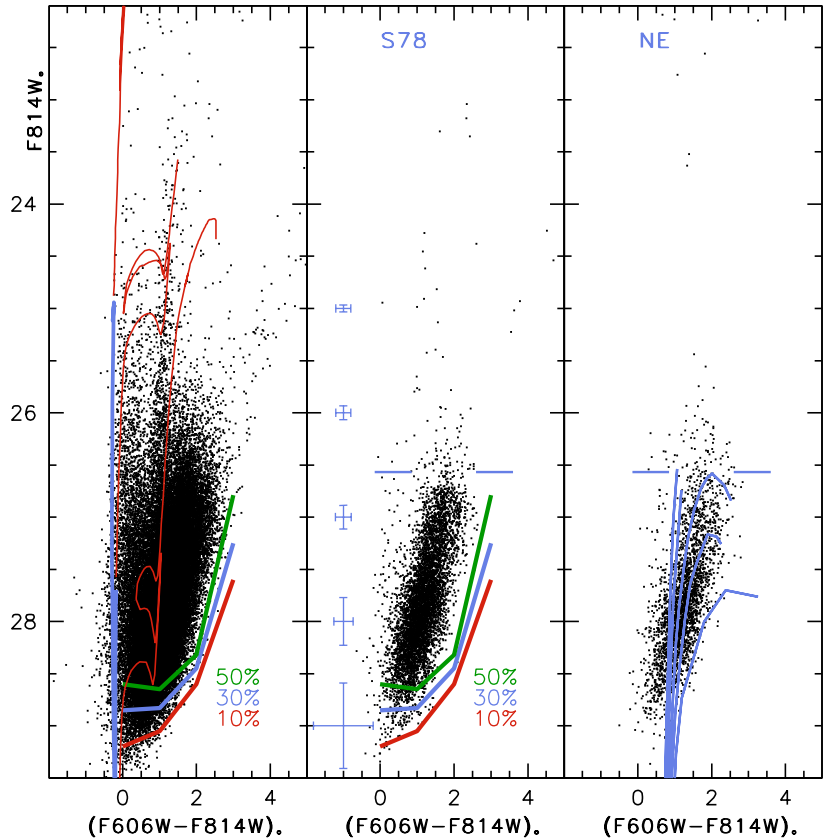} 
\par\end{centering}

\caption{True luminosity vs. true color (obtained by removing only the foreground
extinction and reddening) of stars in different regions of the tidal
tail. The left panel displays the ACS/WFC $F814W_{\circ}$ vs. $(F606W-F814W)_{\circ}$
CMD for the entire ACS field. All stars with absolute \textsc{Sharp}
values less than 0.35 are plotted. Colored lines show the $10\%$
(red), $30\%$ (blue) and $50\%$ (green) completeness levels. The
photometric errors in magnitude and color are indicated by blue crosses
in the center panel. Both are derived from artificial star experiments.
Isochrones from the Padova Library in the VEGAMAG system are plotted
as well, for a metallicity $Z=0.004$ and ages of $8$, $80$, and
$224$ Myr. Post turnoff phases are drawn with a thinner line. The
middle panel displays the CMD for stars in the S78 region, and the
right panel that of stars in the NE quadrant. The luminosity of the
red giant branch tip, as detected in the NE quadrant, is marked by
the horizontal segments in the center and right panels: stars in the
S78 region are fainter and redder due to the presence of dust in that
area. Overplotted in the right panel are fiducial ridge lines of Galactic
globular clusters in the VEGAMAG system (from Bedin et al. \citeyear{bedin_etal05}).
The clusters are (left to right): NGC6341, NGC6752, NGC104 (47Tuc),
NGC5927, NGC6528 whose metallicities are: {[}Fe/H]=$-2.28,-1.56,-0.76,-0.37$
and $-0.04$, respectively. }

\label{f_figura2} 
\end{figure}

The photometric reduction of the final $F606W$ and $F814W$ images
was performed with the \textsc{Daophot/Allstar} package (Stetson \citeyear{stetson87},
\citeyear{stetson94}). The calibration to the ACS VEGAMAG system
was done following the recipes described in Sirianni et al. (\citeyear{sirianni_etal05};
hereafter S05) and Bedin et al. (\citeyear{bedin_etal05}; hereafter
B05). The instrumental magnitudes were transformed both to the HST
VEGAMAG system, and to the Johnson-Cousins system. Magnitudes in the
VEGAMAG system were obtained by adding the zero-points as listed in
B05, and these were dereddened assuming: (i) $E_{B-V}=\EbvSHR$ (see
next section); and (ii) the extinction ratios listed in Table~14
of S05, for a star of spectral type G2. For the stars in common with
the WFPC2 imaging, we found a 3$-\sigma$ clipped mean difference
with the ACS VEGAMAG $I$ magnitudes of $0.003\pm0.005$ mag. No dependence
on magnitude for $I_{ACS}\leq26.5$ was found. This confirms the excellent
consistency of the two scales. To determine the photometric errors
and completeness of the ACS data, we introduced $20000$ artificial
stars into the original drizzled images, and reduced and calibrated
the frames exactly as was done for the original images (see Momany
et al. \citeyear{momany05} for a detailed description and example
of this process). The CMD of $56600$ stars for the whole ACS field-of-view
is shown in the left panel of Fig.~\ref{f_figura2}. The dependence
of completeness on stellar color is summarized by the plotted iso-completeness
lines: we can be confident that in the color range of the RGB all
stars are measured, while only approximately half of any extremely
red stars, if they exist, would be recovered. Nevertheless by looking
at the middle and right panels of Fig.~\ref{f_figura2}, which are
for the areas outside the regions of active star formation, one can
conclude that very few stars with colors $(F606W-F814W)\geq3$ exist
at magnitudes comparable to the RGB tip. These would be carbon stars
along the asymptotic giant branch (see e.g., Fig.~6 in Stetson et
al. \citeyear{stetson_etal98}).

\section{Foreground and internal reddening}

In SHR04 the apparent luminosity of the TRGB was computed by considering
only stars located far from star-forming regions. A few triangular
areas were picked from PC, WF3, and WF4 chips, with the dominant sample
coming from the southern portion of WF4 (see Fig.~2 in that paper).
The WF2 area was excluded since it is within the \ion{H}{1} isodensity
contours, while the WF4 portion is just outside the southernmost contour,
which marks a value of $5\times10^{19}$~cm$^{-2}$ (Fig.~9 in SHR04).
Assuming that internal extinction is negligible in the low gas density
areas, in SHR04 the tip luminosity was then corrected only for the
foreground extinction.

In the case of our new ACS data, we expect that the internal extinction
is negligible for stars located in the north-eastern quadrant of the
ACS field: Fig.~6 of Hibbard et al. (\citeyear{hibbard_etal01})
shows that the \ion{H}{1} density is below the detection limit
in that area. The apparent luminosity of the TRGB was then measured
using stars located in that area, which is also far from SF regions,
hence yielding a nearly reddening-free measurement of the TRGB. On
the other hand, the low stellar density implies a relatively large
uncertainty on the tip position. As a further check, we then repeated
the tip detection using stars in the region with the highest stellar
density, which is that coincident with the S78 object (cf. inset in
Fig.~\ref{f_figura1}). Although this population allows the clearest
detection of the TRGB, at the same time it is affected by high extinction
due to the high gas and dust density ($9\times10^{20}$~cm$^{-2}$),
so for a proper consistency check, we need to estimate the reddening
in the S78 region.

Using the technique explained below, we found that in the dense regions
the observed magnitude of the tip is indeed $\intextinc\pm\errintextinc$~mag
fainter in the F814W band. From S05 we get $A_{F814W}=1.825\times E_{B-V}$
for a G2 spectral energy distribution, so we conclude that the internal
reddening in the S78 region is $E_{B-V}=\intredd\pm\errintredd$.
This is roughly consistent with what we expect based on the gas and
dust column density. The study of Lockman \& Condon (\citeyear{lockman_condon05})
finds $E_{B-V}=a+b\times N_{{\rm HI}}$, with two possible solutions:
$a=0.0017$~mag and $b=1.0\times10^{-22}$~mag~cm$^{2}$, or $a=-0.0073$~mag
and $b=1.5\times10^{-22}$~mag~cm$^{2}$. Inserting the \ion{H}{1}
density of the S78 region in the equations, these yield a reddening
comprised between $E_{B-V}=0.09$ and $E_{B-V}=0.13$. When compared
to the value estimated above, these reddenings suggest that $\sim30\%$
to $\sim100\%$ of the gas and dust could be in front of the stars.
If instead we use the calibration of Bohlin et al. (\citeyear{bohlin_etal78}),
then the reddening is $1.7\times10^{-22}\times N_{{\rm HI}}$~mag~cm$^{2}$,
which gives $E_{B-V}=0.15$, or $\sim40\%$ of the gas and dust in
front of the stars. The foreground reddening to the Antennae is $E_{B-V}=\EbvSHR$
(Schlegel et al. \citeyear{schlegel_etal98}), so the total reddening
in the S78 region is then $E_{B-V}=\EbvS\pm\errEbvS$. {Note
that the color of the RGB in the higher reddening area is not significantly
different from that in the lower reddening area. This is because the
expected shift to redder colors (by ca. $0.1$ mag) is counter balanced
by the shift to the blue due to field crowding (the full results of
the artificial-star experiments will be discussed in the main article
dealing with the SFH of the tidal feature).}

The reddening is also used to calibrate the photometry (see previous
section), so in principle stellar populations of different age and/or
spatial location should be calibrated independently. However, since
in this study we are mainly concerned with the distance, we adopt
only one calibration, which is valid for the old population of the
north-eastern quadrant of the ACS field.

\section{Distance}

We measure the distance of the Antennae using the luminosity of the
TRGB in the $I$-band, a now well established technique (e.g., Rizzi
et al. \citeyear{rizzietal06}, and the references therein). By definition
the distance modulus is $\mu_{0}=(m-M)_{0}=I^{{\rm TRGB}}-M_{I}^{{\rm TRGB}}-A_{I}$,
where $I^{{\rm TRGB}}$ is measured on the luminosity function (LF)
of the RGB population, $M_{I}^{{\rm TRGB}}$ is a function of the
average metallicity of the RGB population, and $A_{I}$ is the extinction
in the $I$-band. To compute the average metallicity one can use a
color-metallicity relation, which gives ${\rm [Fe/H]}$ as a function
of the color of the RGB measured at some pre-defined \emph{absolute}
luminosity of the branch. Therefore one needs $\mu_{0}$, which is
the quantity one is trying to compute. Distance and metallicity are
then computed iteratively, starting from some reasonable guess of
$\mu_{0}$, and working until convergence (see Sakai et al. \citeyear{sakai_etal96}).
The details of this procedure are the following.

\begin{figure}
\begin{centering}
\includegraphics[width=0.7\textwidth]{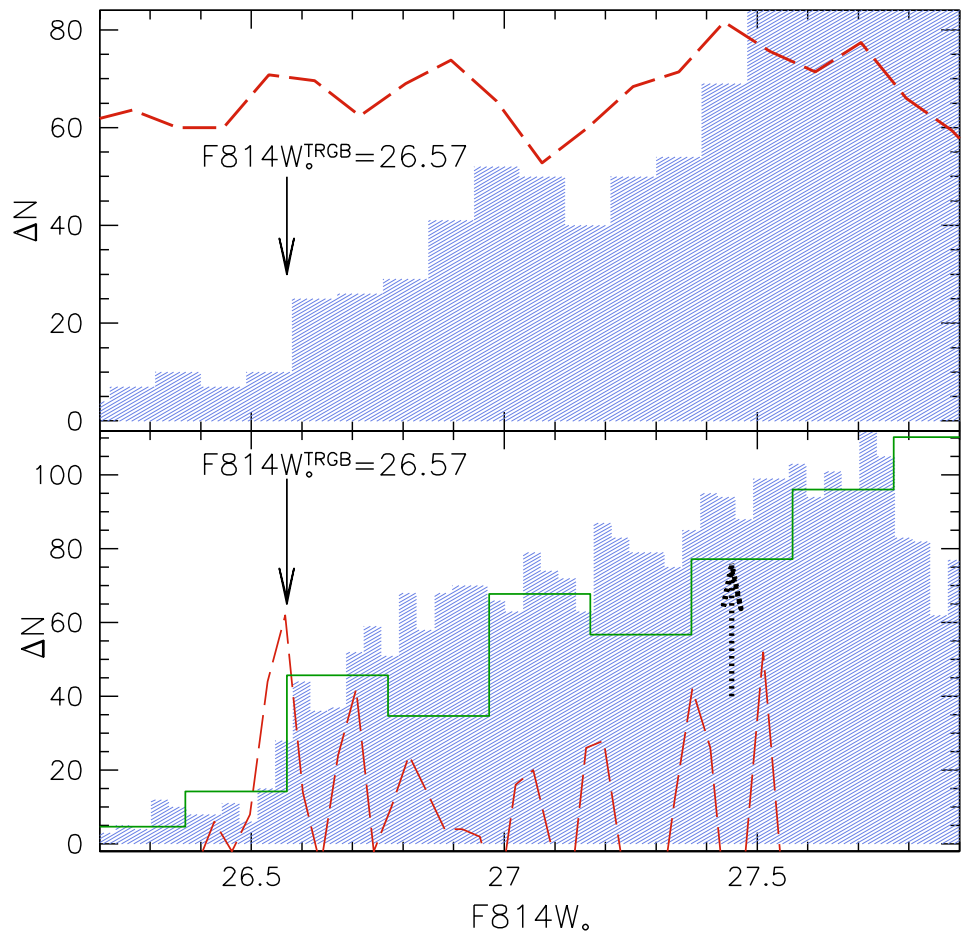} 
\par\end{centering}

\caption{The luminosity functions of the RGB populations from the NE and S78
regions are shown by the shaded histograms in the upper and lower
panel, respectively. The LF of the S78 population has been corrected
for the additional $\intextinc$ magnitudes of internal absorption.
Processing the LFs with a Sobel kernel filter yields the dashed curves,
and in both panels, the arrow marks the RGB tip level, given by the
peak at the highest luminosity of the filtered LFs. If the distance
was approximately $20$ Mpc, as usually assumed in the literature,
then the RGB tip level should be at the luminosity marked by the dotted
arrow, and this is clearly ruled out by our data. The solid histogram
in the lower panel is the LF of the Fornax dwarf spheroidal galaxy
(from Saviane et al. \citeyear{saviane_etal00}), and the agreement
between the LFs is a further hint of the old age of the S78 population.
The discontinuity at $F814W_{0}\approx27.5$ in the LF of the NE population
could be due to the RGB of a younger, $\sim200~{\rm Myr}$ population
(see previous figure). \label{f_tip}}

\end{figure}

To measure the luminosity function (LF) of RGB stars, {RGB
samples were defined by applying two spatial selections, that include
only objects in the S78 region or the NE quadrant away from star-forming
regions} (cf.\ Fig.\ \ref{f_figura1}). The CMDs for these samples
of $2800$ and $6200$ stars (NE quadrant and S78 region, respectively)
are shown in the middle and right panels of Fig.~\ref{f_figura2}.
The RGB of a presumably old (age $\geq$ 2 Gyr) population is clearly
evident in these panels (populations younger than $\sim$2 Gyr do
not generate substantial RGB sequences). This population most likely
arises from the original disk of NGC~4038. In Fig.~\ref{f_tip}
we show the luminosity functions of the RGB populations presented
in the middle and right panels of Fig.~\ref{f_figura2}. The RGB
tip is readily visible and application of a Sobel edge-detection filter
yields $F814W_{\circ}^{{\rm TRGB}}=\tipACSlowred\pm\errtipACSlowred$
for the population of the NE region. Here the error is just the formal
uncertainty of the method, which depends on the bin size. This value
must be corrected for the bias in the observed magnitudes, derived
from the artificial star experiments, which is $0.05$~mag at this
luminosity. The corrected value is then $F814W_{\circ}^{{\rm TRGB}}=\tipACSlowredTrue$~mag.
The TRGB luminosity and color {[}$(F606W-F814W)_{\circ}=\colortipACS$]
were converted from the VEGAMAG system to the Johnson-Cousins system,
yielding $I_{\circ}^{{\rm TRGB}}=\IoTip$. The observed tip luminosity
(i.e., with no bias applied) is $\tipdiffnobias$ magnitudes fainter
than that derived in SHR04. The difference in the two determinations
is probably due to the same photometric bias that was found above
by means of simulations. The $0.05$~mag found here might well turn
into \tipdiffnobias~mag when working near the photometric limit
of the WFPC2 dataset. 

As recalled above, $I_{0}^{{\rm TRGB}}$ can now be converted into
a distance once $M_{I}^{{\rm TRGB}}$ and ${\rm [Fe/H]}$ are computed
iteratively. The metallicity was determined by comparing the color
of the RGB to the fiducial branches of Galactic globular clusters
taken from B05. The luminosities and colors of the fiducials were
computed using values of the distance moduli and reddenings from Harris
(\citeyear{harris96}; catalog version 2003 February). The ${\rm [Fe/H]}$
of the clusters was then plotted against the color of the RGB at $M_{F814W}=-3$
(cf. Da Costa \& Armandroff \citeyear{da90}). The metallicities were
again taken from Harris (\citeyear{harris96}) and a fit was obtained
excluding the two clusters at the metallicity extremes (namely NGC~6341
and NGC~6528): this ensures that a linear relation is a good approximation
to the metallicity--color relation in the range of interest. The mean
color of the RGB stars was then obtained by selecting stars with luminosities
$-3.5<M_{F814W}<-2.5$, and was converted into a mean metallicity
by using the relation obtained above, which has a slope $\Delta{\rm {[Fe/H]}/\Delta{\rm {color}=2.64}}$~dex/mag.
In this way we measure a mean metallicity for the RGB stars of ${\rm {\langle[Fe/H]\rangle}=\feh\pm\errfeh}$.
The error is dominated by the uncertainty on the distance modulus,
and it takes into account the uncertainty on the mean color of the
RGB, that on the photometric zero point and the reddening determination,
and the error on the metallicity-color calibration. 
The value of the mean metallicity is valid only if the RGB stars are
as old as those in Galactic globular clusters. If they are younger,
then since at fixed abundance a younger RGB has bluer colours ({using
isochrones from Girardi et al. (\citeyear{girardi_etal02})} we estimate
that an age decrease of $5$~Gyr changes the color of the RGB at
$M_{F814W}=-3$ by $\sim$0.2~mag) our mean metallicity estimate
is actually a lower limit on the true value. With the metallicity-color
relation quoted above, we can estimate a slope of the metallicity-age
relation of $0.1$~dex/Gyr. We note, however, that Mirabel et al.
(\citeyear{mdl92}) measured a gas-phase oxygen abundance of ${\rm [O/H]\simeq-0.5}$
for the star-forming regions in this vicinity, which must be also
the metallicity of the current generation of stars. Assuming that
the metallicity increases with time, and that oxygen traces iron,
then this value represents an upper limit to the metallicity of the
RGB population. 

To estimate the absolute luminosity of the TRGB, we use the fact that
the mean metallicity of RGB stars computed above is intermediate between
those of $\omega$Cen and 47~Tuc. These two clusters are the fixed
points of the luminosity-metallicity relation established by Bellazzini
et al. (\citeyear{bellazzini_etal04}), with TRGB luminosities set
at $-4.05\pm0.12$~mag and $-3.91\pm0.13$~mag, respectively. Note
that, although it is well-know that $\omega$Cen stars have a range
in metallicity, still more than $80\%$ of the cluster stars share
a common {[}Fe/H] within $0.2$~dex (Suntzeff \& Kraft \citeyear{suntzeff_kraft96},
Norris et al. \citeyear{norris_etal96}, Hilker et al. \citeyear{hilker_richtler00}),
hence the cluster suitability as a TRGB calibrator. Since the metallicity
of Antennae stars is in the scale of Harris (\citeyear{harris96}),
we adopt {[}Fe/H]$=-1.62$~dex for $\omega$Cen and {[}Fe/H]$=-0.79$~dex
for 47~Tuc. A linear interpolation then yields $M_{I}^{TRGB}=\MiTip\pm\errMiTip$.
This luminosity value is valid only for a population with an age comparable
to that of Galactic globular clusters (GGC). Since RGB stars in the
tidal tail were once part of the NGC~4038 disk, it is likely that
they have an age spread. Using theoretical models from Girardi et
al. (\citeyear{girardi_etal02}) we computed the luminosity spread
of the TRGB for ages greater than $5$~Gyr, and metallicities smaller
than ${\rm [Fe/H]=-0.7}$, which amounts to $0.07$~mag ($1\sigma$)
in the $I$ band. On the other hand the same models show that age
and metallicity act in opposite directions, so the likely presence
of an age-metallicity relation will tend to reduce the luminosity
variation, and $0.07$~mag can be taken as an upper limit to the
luminosity spread.

\textit{Our firm distance modulus to the Antennae is then \mbox{(m--M)$_{\circ}$}$=\mMo\pm\emMo$,
corresponding to $\D\pm\eD$~Mpc}. In the error budget we have included
the error in the aperture correction ($0.005$~mag), the zero-point
error of the VEGAMAG photometric system ($0.02$~mag) and the error
on the reddening, the uncertainty in the tip position in the LF ($0.09$~mag),
the error in the HST-Johnson transformation ($0.03$~mag), an uncertainty
of $0.05$~mag to account for a possible age spread, and finally
the error on the absolute magnitude of the tip ($0.13$~mag).

Before comparing our new value of the distance to that obtained in
SHR04, the Population~II distance scale must be discussed. Traditionally
the tip luminosity in the $I$-band was taken as $M_{I}^{{\rm TRGB}}\approx-4.0$
(e.g Da Costa \& Armandroff \citeyear{da90}), with a mild dependence
on the metallicity. However, after Hipparcos released the catalog
of parallaxes for subdwarf stars in the solar vicinity, the traditional
Pop II distance scale was revised. By fitting globular clusters' main
sequences to subdwarf sequences, Carretta et al. (\citeyear{carretta_etal00})
suggested that the so-called `long' distance scale was favored by
the new data. This means that Galactic globular clusters' distance
moduli are on average $\sim$0.2~mag greater than traditional values,
and hence the RGB tip luminosity is greater by the same amount. In
SHR04 such a distance scale was adopted, which led to a distance of
$13.8$~Mpc for the Antennae. However, more recent calibrations of
the Pop II distance scale seem to be moving back to `short' distances.
For example, in the case of 47~Tuc Zoccali et al. (\citeyear{zoccali_etal01})
find $(m-M)_{\circ}=13.27\pm0.14$ by fitting the cluster's white
dwarf (WD) cooling sequence to local WDs with measured parallaxes,
and Bellazzini et al. (\citeyear{bellazzini_etal04}) have recently
published a TRGB luminosity calibration where they adopt $(m-M)_{\circ}=13.31\pm0.14$
for 47~Tuc. The newer distances are then shorter by $\sim0.2$~mag
compared to Carretta et al. (\citeyear{carretta_etal00}), and Gratton
et al. (\citeyear{gratton_etal03}). In this paper we adopt the more
recent distance scale, but if we adopted the tip luminosity of SHR04,
the distance of the Antennae would be $\Dlong$~Mpc, fully compatible
with our previously derived value of $\Dold\pm\errDold$~Mpc.

Note that, although the distance we measure strictly applies only
to stars in the southern tail of the Antennae, it can be safely used
for the whole system. In fact adopting the distance of this paper,
the numerical model that was matched to the \ion{H}{1} kinematics
in Hibbard et al. (\citeyear{hibbard_etal01}), gives a maximum tidal
extent (tip to tip, in real space) of $90$~kpc. The maximum line-of-sight
depth is $36$~kpc (tip of N tail to base of N disk). The maximum
line-of-sight depth between our pointing and anywhere else in the
system is $16$~kpc (tip of southern tail to base of N disk). If
we adopt $100$~kpc as an upper limit to the maximum line-of-sight
depth, then the distance modulus to any other region of the Antennae
will be less than $0.02$~mag different from the one we measure.

For distances of the order of $13$~Mpc, the recession velocity for
the Antennae differs from that inferred from the flow model of Tonry
et al. (\citeyear{tonry_etal00}) by approximately $500\kms$. The
model predicts a peculiar velocity dispersion of $187\kms$, so this
difference from the model is not extreme, being at the level of $2.7~\sigma$.
We note also that a similar difference is present for another galaxy
near the position of the Antennae; in Fig.~11 of Tonry et al. (\citeyear{tonry_etal00})
there is an object near the Antennae that has a radial velocity $\sim300$~$\kms$
larger than the model prediction. Moreover, objects near the Antennae
are beginning to fall within the gravitational influence of the Great
Attractor (GA), and the quoted figure in Tonry et al. (\citeyear{tonry_etal00})
shows several other objects with large residual velocities pointing
toward the GA\@. Thus we believe our new distance does not generate
any significant discrepancies with the observed recession velocity
of the Antennae system.

\section{Discussion}

With our new HST/ACS imaging of the tidal tail region studied in SHR04,
more than two magnitudes of the red-giant branch of an old (age $\geq$
2 Gyr) population is readily visible. The resulting clear detection
of the red-giant branch tip confirms the short distance found in the
earlier paper. The downward revision of the distance to the Antennae
has far reaching scientific implications: some of these are discussed
in SHR04, and an update is given here.

\subsection{Young massive clusters}

An immediate implication is that the linear scales of, e.g., Antennae
clusters are reduced by a considerable factor, which depends on the
distance adopted in the original study. For example, using two different
techniques Whitmore et al. (\citeyear{whitmore_etal99}) estimated
a median effective radius of \textit{$r_{{\rm eff}}=4\pm1$}\textit{\emph{~pc
and}} \textit{$4.6\pm0.4$}\textit{\emph{~pc for the young clusters.
This is larger than the median $r_{{\rm eff}}$ of globular clusters
in the Milky Way ($\sim3$~pc, see van den Bergh \citeyear{vandenbergh96}),
and of clusters in nearby starburst galaxies ($2$--$3$~pc, see}}
\textit{\emph{Meurer et al.}} \textit{\emph{\citeyear{meurer_etal95}).
With the shorter distance all linear dimensions are reduced by a factor
$1.4$, so the radii quoted above become $2.8$~pc and $3.2$~pc,
respectively. Therefore the size of the clusters}} formed as a consequence
of the Antennae \textit{\emph{merger now agree with those of well-studied
objects in the nearby Universe, and with}} \textit{$r_{{\rm eff}}$~$\approx3$--$6$~}\textit{\emph{pc
for clusters created in other mergers}} \textit{\emph{(Schweizer et
al. \citeyear{schweizer_etal96};}} \textit{\emph{Miller et al.}}
\textit{\emph{\citeyear{miller_etal97};}} \textit{\emph{Whitmore
et al.}} \textit{\emph{\citeyear{whitmore_etal97};}} \textit{\emph{Carlson
et al.}} \textit{\emph{\citeyear{carlson_etal98}).}}

Moreover all luminosities are reduced by a factor of $\sim2$ compared
to the most commonly quoted distance of $19.2$~Mpc. Thus, along
with the reduction factor of $1.4$ in linear scales, this implies
a smaller projected length for the tails while the radial velocities
would remain unchanged, leading to a shorter timescale for the interaction
(Hibbard et al. \citeyear{hibbard_etal01}). The luminous to dynamical
mass estimate for the tidal dwarf candidates (Hibbard et al. \citeyear{hibbard_etal01})
would be $1.4$ times lower ($M_{{\rm lum}}/M_{{\rm dyn}}\sim0.2-0.5$),
implying a more significant dark-matter content, as recently inferred
for other tidal dwarfs (Bournaud et al. \citeyear{bournaud_etal07}).

Observations of advanced stages of mergers has lead to the idea that
the end product of a disk-disk encounter could be a giant elliptical
(E) galaxy (Toomre \& Toomre \citeyear{tt72}; Toomre \citeyear{toomre77};
Schweizer \citeyear{schweizer87}). However there are two problems
with this idea, both related to the presence of rich globular cluster
(GC) populations around E galaxies. The first problem is that E galaxies
have higher specific frequencies $S_{N}$ of GCs (number of clusters
per unit luminosity, see Harris \& van den Bergh \citeyear{harris_vdb81})
compared to spiral galaxies (van den Bergh \citeyear{vandenbergh82}).
While massive disk galaxies have $S_{N}\sim1$, Es have $S_{N}\sim2$--$5$
depending on the environment (e.g., Brodie \& Strader \citeyear{brodie_strader06}).
The second problem is that the mass function (MF) of GCs is log-normal
(i.e., Gaussian when plotted vs. the logarithm of the mass), while
young massive clusters show power-law MFs with exponents close to
$\alpha=-2$ (Elmegreen \& Efremov \citeyear{elmegreen_efremov97},
Hunter et al. \citeyear{hunter_etal03}, Bik et al. \citeyear{bik_etal03},
De Grijs et al. \citeyear{degrijs_etal03}, Zhang \& Fall \citeyear{zhang_fall99}).
These two issues are discussed more thoroughly in the following sections.

\subsubsection{Specific frequency}

The fact that new clusters are formed in mergers, as first shown by
\emph{HST} observations of NGC~1275 (Holtzman et al. \citeyear{holtzman_etal92}),
suggests that they could increase $S_{N}$ as long as new clusters
are formed with higher efficiency with respect to field stars, compared
to that of the primordial cluster population of the interacting disks.
In the case of the Antennae, Whitmore \& Schweizer (\citeyear{whitmore_schweizer95})
estimated that $N_{{\rm t}}>700$ new clusters have been formed in
the interaction, and that in a Hubble time the system will fade from
$M_{V}=-23$ to $M_{V}=-21.5$. By definition $S_{N}=N_{{\rm t}}\,10^{0.4(M_{V}+15)}$,
so with the above parameters $S_{N}=1.8$ for the system. Our new
distance modulus is $1.7$ magnitudes shorter than $(m-M)_{0}=32.3$
adopted by Whitmore \& Schweizer (\citeyear{whitmore_schweizer95}),
so with a fainter $M_{V}=-19.8$, the frequency becomes $S_{N}=8.4$.
This is comparable even to Es with the richest cluster systems, so
it might appear that $S_{N}$ is not a problem in the case of the
Antennae. However it is difficult to attach an uncertainty to this
number. On one hand, a substantial number of clusters is below the
detection limit, and more clusters will be formed in the future perigalactic
passages, so $S_{N}$ seems underestimated. But on the other hand
a large fraction of clusters will be destroyed by internal and external
dynamical processes (see e.g., Bastian \& Gieles \citeyear{bastian_gieles07}),
and this will drive $S_{N}$ down. To gain some insight we can look
at merger remnants, which should represent the future evolution of
the Antennae. In the case of NGC~7252 Miller et al. (\citeyear{miller_etal97})
find $S_{N}=2.5$ after $15$~Gyr of fading, and a post-fading $S_{N}=2.9$
is computed in Schweizer et al. (\citeyear{schweizer_etal96}) for
NGC~3921. So it is possible that the net effect of the cluster evolution
will be to reduce the $S_{N}$ of the Antennae, but thanks to its
presently high value, the final value might well still be compatible
with that of a field elliptical galaxy. {Note that about two
thirds of the clusters should be destroyed, which is compatible with
the results of Vesperini (\citeyear{vesperini00}).}

\subsubsection{Mass function}

It is been observed that the LF of young-massive clusters (YMC) can
be represented by two power-laws of different slope (Whitmore et al.
\citeyear{whitmore_etal99}), and with a bend at $M_{V}\approx-10.4$.
This is discussed in the scenario of the possible transformation of
an initial power-law MF into a log-normal MF via destruction of the
lower mass clusters. Adopting $10$~Myr for the age, and using solar
metallicity models from Bruzual \& Charlot \citeyear{bc93}, they
find that the bend corresponds to $10^{5}\,{\rm M}_{\odot}$. Although
the peak of the MF for GCs in the Milky Way is at a higher mass of
$\approx2\times10^{5}\,{\rm M}_{\odot}$, Whitmore et al. (\citeyear{whitmore_etal99})
claim that, given the uncertainties on the location of the bend and
on the $M/L$ ratio given by the models, the two mass values are compatible.
Therefore the proposal is that selective destruction of clusters less
massive than the peak should eventually lead to a GC-compatible, log-normal
MF, as proposed by Fall \& Zhang (\citeyear{fall_zhang01}). With
a shorter distance, the change of slope of the broken power-law of
Whitmore et al. (\citeyear{whitmore_etal99}) is at $\sim\newMassWetal\, M_{\odot}$,
which makes its identification with the MF peak of old Galactic GCs
less likely. Moreover Zhang \& Fall (\citeyear{zhang_fall99}), modeling
the age distribution of YMCs find that the MF is a power-law of $\alpha=-2$,
with no signs of a break. They interpret the bend in the LF as an
effect of a MF truncated at $10^{6}M_{\odot}$, and of the fading
of the more massive clusters. Note also that cluster destruction would
leave behind a power-law LF with $\alpha=-2$ in the brightest part,
and a shallower LF in the fainter part, while Whitmore et al. (\citeyear{whitmore_etal99})
find that $\alpha=-2$ in the \emph{fainter} part, and a \emph{steeper}
LF in the brightest part, which can indeed be interpreted as an effect
of fading (Gieles et al. \citeyear{gieles_etal06}).

If the MF is a power-law, then a change of distance only affects the
mass of the most massive clusters generated by the merger. Assuming
a distance of $\DWSninefive$~Mpc Whitmore \& Schweizer (\citeyear{whitmore_schweizer95})
found that the cluster luminosity function ends at $M_{V}\simeq\MvWSninefive$,
which translates into a mass of $\simeq10^{7}{\rm M_{\odot}}$. Old
Galactic GCs have average logarithmic mass $<\log(M)>=5.3$ and a
dispersion $\sigma_{\log{\rm M}}=0.49{\rm \, dex}$ (e.g., Brodie
\& Strader \citeyear{brodie_strader06}), so the largest clusters
in the Antennae would be $3.5\sigma$ larger than the (logarithmic)
average mass of Galactic GCs, and the most massive young clusters
in the Antennae would be almost twice as massive than their counterparts
in the Milky Way. Since massive clusters are not destroyed, this would
pose another problem for the merger-induced formation of globular
clusters. Instead with the new distance the most massive clusters
in the Antennae have masses of $\approx2\times10^{6}{\rm M_{\odot}}$,
which is well within the mass limits of Galactic GCs.

Although the models of Fall \& Zhang (\citeyear{fall_zhang01}) can
transform a power-law MF into a log-normal one, this happens only
with a restricted set of parameters. Instead the simulations of Vesperini
(\citeyear{vesperini00}) have shown that an initial log-normal MF
stays log-normal for large variations of the initial conditions, and
the average mass and dispersion tend to reach those of old GCs after
a few Gyr of dynamical evolution. This is despite the fact that $\sim50\%$
of the clusters are destroyed in the process. This convergence would
explain why the GC MF is almost universal. In this respect, some investigations
have warned that log-normal and power-law mass (or luminosity) functions
are significantly different only at the faint end, where they are
usually hard to determine due to incompleteness of the data (e.g.,
Anders et al. \citeyear{anders_etal07}). The Antennae are a particularly
important case, since their YMC population is one of the youngest
(only a few tens of Myr old), so it can tell us what is the shape
of the MF of merger-induced clusters at the very beginning. With a
different analysis of the \emph{HST} imaging presented in Whitmore
et al. (\citeyear{whitmore_etal99}), a new LF was obtained by Anders
et al. (\citeyear{anders_etal07}) and converted into a MF assuming
an average age of $25$~Myr and solar metallicity. They found that
a log-normal function provides a better fit, with $<\log(M)>=4.2$
and $\sigma_{M}=0.85$ dex (and assuming again $19.2$~Mpc distance).
With a shorter distance the average mass of Anders et al. (\citeyear{anders_etal07})
becomes $<\log(M/M_{\odot})>=3.9$, so more dynamical evolution is
required to shift it to a value comparable to old GCs. While an initial
log-normal MF would be appealing, preliminary results, based on new
and deeper HST/ACS imaging, seem to show that the MF is a power-law
even past the older detection threshold (Whitmore \citeyear{whitmore06}).
So the issue of GC creation in mergers is still open. Either conditions
in the primordial Universe made clusters with a characteristic mass,
and today they do not. Or clusters are always born with power-law
MFs which evolve to log-normal. A definite answer will need more realistic
simulations of cluster evolution in a time-variable potential like
that of a merger. Note also that evolution of the cluster MF has perhaps
been detected in intermediate-age mergers like NGC~1316 and NGC~3610
(Goudfrooij et al. \citeyear{goudfrooij_etal04}; Goudfrooij et al.
\citeyear{goudfrooij_etal07}).

\subsection{SN 2004gt}

SN 2004gt is of great interest because its presence in the Antennae
may permit the strongest yet constraints on the progenitor of Type
Ic SNe. It is currently debated whether the progenitors of these SNe
are high-mass single stars or intermediate-mass stars in a binary
system (see below), and since a small fraction of type Ic SNe are
associated with gamma-ray bursts (GRB, e.g., Bloom et al. \citeyear{bloom_etal99};
Stanek et al. \citeyear{stanek_etal03}; Hjorth et al. \citeyear{hjorth_etal03};
Malesani et al. \citeyear{malesani_etal04}; Pian et al. \citeyear{pian_etal06}),
being able to discriminate between these two hypothesis is of great
interest for GRB models as well. Indeed no progenitor of a type Ic
SN has ever been found, and we only have upper luminosity limits for
five stars (see Table~1 in Maund et al. \citeyear{maund_etal05}).
Although SNe of type Ic have been identified in galaxies closer than
the Antennae, the depth of \emph{HST} archival images permits to reach
fainter absolute magnitudes. Nevertheless both Maund et al. (\citeyear{maund_etal05})
and Gal-Yam et al. (\citeyear{gal-yam_etal05}) failed to find the
progenitor of SN2004gt, and only upper limits could be established.
Still these limits are fainter than all previous determinations, and
consequently place the strongest constraints on possible candidates.

The first study adopted a conventional $19.2$~Mpc distance, while
the second one used the shorter $13.8$~Mpc distance established
in SHR04. With the distance proposed in this paper (which is only
$<4\%$ smaller than our previous one) the results of Maund et al.
(\citeyear{maund_etal05}) can be put in better agreement with those
of Gal-Yam et al. (\citeyear{gal-yam_etal05}), as we now show. The
conclusion of Gal-Yam et al. (\citeyear{gal-yam_etal05}) is that
the progenitor of SN2004gt, if a Wolf-Rayet (WR) star, must have belonged
to one of the more evolved types. This is because as a WR star evolves
and its envelope gets progressively stripped, it becomes hotter and
fainter (see, e.g., the tracks plotted in Maund et al. \citeyear{maund_etal05}),
and all progenitors brighter than $1.25\times10^{4}\, L_{\odot}$
in the $V$ band ($M_{V}>-5.2$) are excluded by Gal-Yam et al. (\citeyear{gal-yam_etal05}).
So the progenitor of SN2004gt must have been of type WNE (where the
products of hydrogen-burning through CNO-cycle are exposed), or WC/WO
(where the products of helium burning are exposed). See Maeder \&
Conti (\citeyear{maeder_conti94}) for more details.

The approach of Maund et al. (\citeyear{maund_etal05}) was to perform
a direct comparison with stellar tracks from the Geneva group, computed
with a prescription for mass loss. The magnitude and color limits
of their photometry were converted into a limit in bolometric luminosity
dependent on the effective temperature, and all stars having their
end-point evolution fainter than that limit were accepted as viable
progenitors. The permitted Hertzsprung-Russell diagram region is that
of relatively faint and blue stars, so all red supergiants are excluded,
in agreement with the result of Gal-Yam et al. (\citeyear{gal-yam_etal05})
illustrated above. The end points of $60$ and $85{\rm M_{\odot}}$
stars do reach this region of luminosities smaller than $\sim10^{5}{\rm L_{\odot}}$
and temperatures greater than $\sim30,000{\rm K}$ while $40{\rm M_{\odot}}$
and $120{\rm M_{\odot}}$ tracks are always too luminous. Adopting
$13.3$~Mpc, all luminosities become $0.3\,{\rm dex}$ smaller, and
all WR evolutionary end-points are above the detection threshold,
and thus in principle should be excluded. However WR stars fainter
than the limit computed by Maund et al. (\citeyear{maund_etal05})
do exist (see Vacca \& Torres-Dodgen \citeyear{vacca_torres-dodgen90}),
so our shorter distance reveals an inconsistency in one of the thresholds
computed by Maund et al. (\citeyear{maund_etal05}). Indeed the luminosity
limits for the hottest stars are more uncertain due to the particular
approximation for the WR spectral energy distribution that was adopted,
and to the uncertainties on the bolometric corrections (Maund, priv.
comm.). So at $13.3$~Mpc progenitors of $40\,{\rm M}_{\odot}$ could
still be possible.

The general conclusion is then that both Maund et al. (\citeyear{maund_etal05})
and Gal-Yam et al. (\citeyear{gal-yam_etal05}) support a type Ic
progenitor which is a WR star of the most evolved types, and with
a main-sequence mass in the lowest range of these wind-dominated stars
(see, e.g., Maeder \& Conti \citeyear{maeder_conti94} for a review).
However, an alternative scenario for SNe Ic is an origin from a star
of lower mass than a WR, and in a binary system. In that case the
envelope is stripped through interaction with its companion (see Pods
et al. \citeyear{pods_etal04} and references therein). This possibility
cannot be constrained by the SN in the Antennae, so this issue is
still open.

\subsection{Star formation induced by the merger}

The Antennae is often used as an example of the kind of `violent'
processes occurring in mergers. For example, Sanders \& Mirabel (\citeyear{sanders_mirabel96})
include this system in their compilation of Luminous Infrared Galaxies,
which owe their high IR luminosity to a starburst. This is a class
of galaxies whose total IR luminosity exceeds $10^{11}\, L_{\odot}$:
according to Vigroux et al. (\citeyear{vigroux_etal96}), who assumed
a distance of $20$~Mpc, this is precisely the IR luminosity of the
Antennae. Since the total IR luminosity is now $\sim\newLumIR\, L_{\odot}$,
the Antennae should be classified in the regime of normal galaxies.
This is perhaps not surprising, since most interacting galaxies have
normal IR luminosities (Bushouse et al. \citeyear{bushouse_etal88}).
And in particular the Antennae are the object opening the so-called
Toomre sequence (Toomre \citeyear{toomre77}), so the star-formation
rate (SFR) is not expected to be that of the more advanced stages.
Indeed Charmandaris et al. (\citeyear{charmandaris_etal00}) studied
the SFR along a modified Toomre sequence, as traced by the $15$~$\mu{\rm m}$
to $7$~$\mu{\rm m}$ flux ratio, and found that the ratio is small
for galaxies in early stages of interaction, it increases by a factor
of $\sim5$ in merging/starburst systems, and it goes back to pre-starburst
values in merger remnants. Although the Antennae was not part of the
sequence studied by Charmandaris et al. (\citeyear{charmandaris_etal00}),
the low IR luminosity deduced here is in line with the low SFR expected
for a merger in its earliest stages.

Another, indirect confirmation of this is the fact that, thanks to
the shorter distance, the luminosity function of the X-ray sources
(XLF) in the Antennae becomes comparable with that of M82. The most
recent XLF is presented in Zezas \& Fabbiano (\citeyear{zezas_fabbiano07}),
and it is a power-law reaching maximum luminosities of $\log L_{X}^{{\rm max}}\simeq39.8\,{\rm erg\, sec^{-1}}$
in the $0.1$--$10.0~{\rm keV}$ band, for an assumed $19~{\rm Mp}$c
distance. The maximum luminosity becomes $\log L_{X}^{{\rm max}}\simeq39.5\,{\rm erg\, sec^{-1}}$
after subtracting $0.3~{\rm dex}$, which is the factor of $2$ reduction
in luminosity implied by our shorter distance. This means that $L_{X}^{{\rm max}}$
is close to that of the XLF of M82 (e.g., Zezas \& Fabbiano \citeyear{zezas_fabbiano02}),
which also has a similar slope. The high-luminosity tail of the XLF
is populated by high-mass X-ray binaries, which are a signature of
a young stellar population. The similarity of the XLFs of the Antennae
and M82 means that the underlying stellar population has comparable
age, i.e. that the strength of the on-going SF is also comparable.
The SFR per unit area of M82 is comparable to that of NGC~520 (Kennicutt
\citeyear{kennicutt98}), which is classified as pre-starburst in
the modified Toomre sequence of Charmandaris et al. (\citeyear{charmandaris_etal00}).
So we might infer that the SFR in the Antennae is also comparable
to that of NGC~520 and M82, in turn a sign of an early merger stage.
In fact another consequence of the reduced distance is that the number
of so-called ultra-luminous X-ray sources (ULX; those exceeding $10^{39}$~erg~cm$^{-2}$~sec$^{-1}$)
become comparable to the number of ULXs found in NGC~520. Eighteen
ULXs were found by Zezas \& Fabbiano (\citeyear{zezas_fabbiano02})
in the Antennae, for a distance of $29$~Mpc. With our distance,
these become \nULX\, which is a number closer to the three--four
ULX sources detected by Read et al. (\citeyear{read05}) in NGC~520.
The maximum X-ray luminosities quoted above are also comfortably small:
indeed Zezas \& Fabbiano (\citeyear{zezas_fabbiano07}) find that
even luminosities up to $10^{40}$~erg~cm$^{-2}$~sec$^{-1}$ can
be radiated by a black-hole of $\sim80~M_{\odot}$, which in turn
can be the end-point of normal stellar evolution in a binary system.
In fact Zezas \& Fabbiano (\citeyear{zezas_fabbiano07}) propose to
move the limit of ULX sources above $\log L_{X}=40$, since only above
such luminosities a non-standard scenario is required (like an intermediate-mass
black hole or beamed radiation).

In summary, with the shorter distance determined here, the Antennae
remain a spectacular system, but it is no longer an extreme merger
in terms of its consequences (star-burst luminosity, cluster population,
X-ray sources, etc). A full appreciation of the properties of this
system in light of our new firm distance will have significant implications
for many areas of Astronomy.

\acknowledgements
This paper has been completed thanks to a stay of Y.M. at ESO/Chile
(through a Director General grant to I.S.) and a stay of I.S. at Mt
Stromlo Observatory, supported both by ESO (through the same DG grant)
and the Australia National University (grant ARC DP0343156). We thank
Justyn Maund and Mark Gieles for constructive discussions about SN 2004
gt and young stellar clusters in the Antennae. The National Radio
Astronomy Observatory is a facility of the National Science Foundation
operated under cooperative agreement by Associated Universities, Inc.

\clearpage{} %

\clearpage{} 

\clearpage{} 

\end{document}